\documentclass[
 twocolumn,
 groupedaddress,
 showpacs,
 showkeys,
 preprintnumbers,
 amsmath,amssymb,
 aps,
 pra,
 longbibliography,
 floatfix
 ]{revtex4-1}
 
\usepackage[normalem]{ulem}
\usepackage{tikz}
\usepackage{graphicx}
\usepackage[colorlinks=true,citecolor=blue,urlcolor=blue]{hyperref}

\def\endproof{\vrule height6pt width6pt depth0pt}

\begin{document}



\title{Minimal true-implies-false and true-implies-true sets of propositions in noncontextual hidden-variable theories}


\author{Ad\'an Cabello}
 \email{adan@us.es}
 \affiliation{Departamento de F\'{\i}sica Aplicada II, Universidad de Sevilla, E-41012 Sevilla, Spain}

\author{Jos\'e R. Portillo}
 \email{josera@us.es}
 \affiliation{Departamento de Matem\'atica Aplicada I, Universidad de Sevilla, E-41012 Sevilla, Spain}
 \affiliation{Instituto Universitario de Investigaci\'on de Matem\'aticas de la Universidad de Sevilla (IMUS), E-41012 Sevilla, Spain}

\author{Alberto Sol\'{\i}s}
 \email{solisencina@us.es}
\affiliation{Departamento de Matem\'atica Aplicada I, Universidad de Sevilla, E-41012 Sevilla, Spain}

\author{Karl Svozil}
 \email{svozil@tuwien.ac.at}
 \affiliation{Institute for Theoretical Physics, University of Technology Vienna, Wiedner Hauptstrasse 8-10/136, 1040 Vienna, Austria}


\date{\today}



\begin{abstract}
An essential ingredient in many examples of the conflict between quantum theory and noncontextual hidden variables (e.g., the proof of the Kochen-Specker theorem and Hardy's proof of Bell's theorem) is a set of atomic propositions about the outcomes of ideal measurements such that, when outcome noncontextuality is assumed, if proposition $A$ is true, then, due to exclusiveness and completeness, a nonexclusive proposition $B$ ($C$) must be false (true). We call such a set a {\em true-implies-false set} (TIFS) [{\em true-implies-true set} (TITS)]. Here we identify all the minimal TIFSs and TITSs in every dimension $d \ge 3$, i.e., the sets of each type having the smallest number of propositions. These sets are important because each of them leads to a proof of impossibility of noncontextual hidden variables and corresponds to a simple situation with quantum vs classical advantage. Moreover, the methods developed to identify them may be helpful to solve some open problems regarding minimal Kochen-Specker sets.
\end{abstract}


\maketitle


\section{Introduction}


The assumption of outcome noncontextuality is the assumption that ideal measurements reveal preexisting noncontextual outcomes. Kochen and Specker \cite{Specker60,KS65,KS67} and Bell \cite{Bell66} proved that there is a conflict between outcome noncontextuality and quantum theory (QT). They pointed out that, for dimension $d \geq 3$, there are sets of atomic propositions (represented in QT by rays in a $d$-dimensional Hilbert space) that do not admit an assignment of noncontextual outcomes once we make the following extra assumptions. (i) {\em Exclusiveness}: exclusive propositions (represented in QT by orthogonal rays) cannot both be assigned the value true. (ii) {\em Completeness}: complete sets of exclusive propositions (represented in QT by $d$ mutually orthogonal rays) cannot all be assigned the value false. These sets are called Kochen-Specker (KS) sets.

It was later pointed out that the conflict between outcome noncontextuality and QT occurs even without assumptions (i) and (ii). Instead, for some linear combinations of correlations, the assumption of outcome noncontextuality, by itself, establishes limits that are violated by QT \cite{KCBS08,Cabello08}. These limits are called noncontextuality (NC) inequalities. NC inequalities generalize Bell inequalities \cite{Bell64} to scenarios where measurements cannot be distributed between separated parties. The quantum violation of some NC inequalities reveal that the conflict also occurs for single particles prepared in arbitrary quantum states \cite{Cabello08,BBCP09,YO12,KBLGC12}. From the perspective of NC inequalities, KS sets are a particular type of contextuality sets, defined as sets of observables for which outcome noncontextuality contradicts the quantum predictions. KS sets can be converted into NC inequalities whose violation reveals quantum state-independent contextuality \cite{KZGKGCBR09, ARBC09}, into Bell inequalities with quantum violation saturating the nonsignaling bound \cite{Cabello01,AGA12}, and into proofs of nonlocality via local contextuality \cite{Cabello10,CAB12,LHC16}.

Every linear combination of correlations appearing in a NC inequality can be expressed as a positive linear combination of probabilities of events or propositions and represented by a graph, called a graph of exclusivity, in which exclusive propositions are represented by adjacent vertices. It was later found \cite{CSW10,CSW14,CDLP13} that QT violates a NC or Bell inequality written this way if and only if its corresponding graph of exclusivity is imperfect, i.e., contains, as induced subgraphs, odd cycles of length five or more (i.e., pentagons, heptagons, etc.), or their complements. Therefore, every proof of contextuality (i.e., impossibility of assigning preexisting noncontextual outcomes to ideal measurements) can be associated to an imperfect graph. This includes any proof, with or without inequalities, of the KS \cite{Specker60,KS65,KS67} and Bell \cite{Bell66} theorems. Reciprocally, every imperfect graph can be used to prove that QT cannot be explained with noncontextual hidden-variable theories \cite{CSW10,CSW14}.

Interestingly, the sets of propositions represented by some specific imperfect graphs allow us to present the conflict between QT and hidden variables in a very appealing way, namely, by pointing out a contradiction between QT and a prediction with certainty of the noncontextual hidden-variable theory. Proofs of this type have been presented by Stairs \cite{Stairs83}, Hardy \cite{Hardy93}, and others \cite{Clifton93a, Clifton93b,Clifton93c,CG95,CT11,CBTB13}. In addition, these imperfect graphs play a fundamental role in the proofs of the KS theorem of Bell \cite{Bell66} and Kochen and Specker \cite{KS67} and in some other proofs of quantum state-independent contextuality \cite{YO12,CKP16}. The purpose of this paper is to identify the minimal (i.e., having the smallest set of vertices) of these imperfect graphs for any dimension and explain how they are related to previous proofs of impossibility of noncontextual hidden variables.

Hereafter, by atomic propositions we will mean statements the form ``outcomes $o_1$ and $o_2$ will be respectively obtained when observables $O_1$ and $O_2$ will be jointly measured on the same physical system,'' where $O_1$ and $O_2$ are assumed to be observables represented in QT by rank-one projectors that commute. Each atomic proposition is represented in QT by a ray in a Hilbert space. Two propositions are exclusive when both cannot be simultaneously true. Exclusive propositions are represented in QT by mutually orthogonal rays. A set of mutually exclusive propositions constitutes a context. A context is complete when one of the propositions must be true.
Greechie orthogonality diagrams~\cite{Greechie71} provide a convenient way to represent the graphs of exclusivity, as they represent contexts as single smooth lines (such as circles or straight unbroken lines) connecting mutually (atomic) exclusive propositions, which are represented as small circles; contexts intertwining at a single proposition are represented as nonsmoothly connected lines, broken at that proposition. For better readability non-intertwining propositions belonging to just one context are not depicted. The assumption of outcome noncontextuality assigns the same truth value (true or false) to any proposition with independence of the context.


\section{True-implies-false and true-implies-true sets}

\begin{figure}
	\begin{center}
		\setlength{\tabcolsep}{0em}
		\begin{tabular}{cc}
			\includegraphics{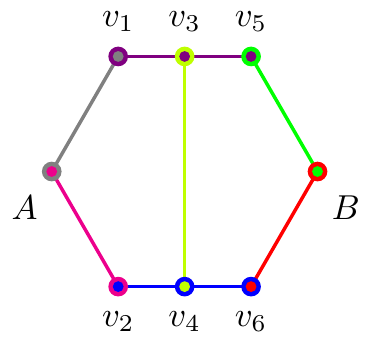} & \includegraphics{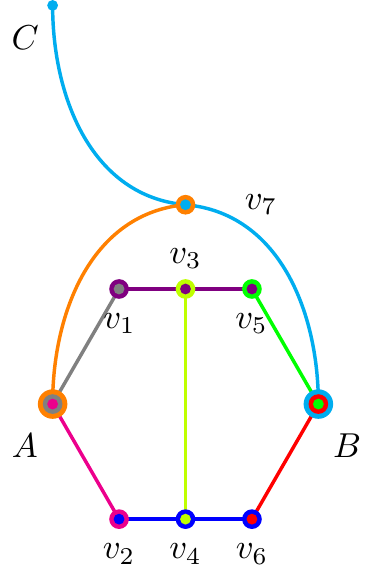} \\
			(a) & (b)
		\end{tabular}
	\end{center}
	\caption{\label{Fig1}
		(Color online)
		Greechie orthogonality diagrams of the minimal
		(a)~TIFS and (b)~TITS in $d=3$.
		Small circles represent propositions, smooth lines represent complete sets
		(i.e., sets in which one and only one of the propositions must be true); in particular, they indicate
		that any pair of propositions connected by a smooth line cannot both be true (exclusiveness).
		(a)~If $A$ is true then $B$ is false~\cite{KS65}.
		(b)~If $A$ is true then $C$ is true~\cite{KS67}.
		These sets are realizable in $S^2$ by taking,
		for instance,
		$v_A = ({1,1,1} )/\sqrt{3}$,
		$v_1 = ({1,-1,0} )/\sqrt{2}$,
		$v_2 = ({1,0,-1} )/\sqrt{2}$,
		$v_3 = ({0,0,1} ) $,
		$v_4 = ({0,1,0} )$,
		$v_5 = ({1,1,0} )/\sqrt{2}$,
		$v_6 = ({1,0,1} )/\sqrt{2}$,
		$v_B = ({-1,1,1}) / \sqrt{3} $,
		$v_7 = ({0,1,-1} )/\sqrt{2}$, and
		$C = ({2,1,1} )/\sqrt{6} $.
		In QT, the proposition $v_i$ is represented by the projector $| v_i \rangle\langle v_i |$.
		To obtain a TIFS or a TITS in $d=4$ it
		is enough to add $\langle v| = (0,0,0,1)$, and similarly to obtain TIFS or TITSs in higher dimensions \cite{CG96}.}
\end{figure}

We define a true-implies-false set (TIFS) [true-implies-true set (TITS)] as a set $S$ ($S'$) of propositions represented in QT by rays in a Hilbert space such that, when outcome noncontextuality is assumed, due to exclusiveness and completeness, if proposition $A \in S$ is true,
then a nonexclusive proposition $B \in S$ must be false (a nonexclusive proposition $C \in S'$ must be true). Explicit examples of a TIFS and a TITS are shown in Figs.~\ref{Fig1}(a) and~\ref{Fig1}(b), respectively. A TIFS or TITS is said to be critical if the set resulting from removing any element is not a TIFS or TITS, respectively. A TIFS or TITS in dimension $d$ is said to be minimal if there are not TIFS or TITS, respectively, with less propositions in dimension $d$.


Any TIFS or TITS, by itself, constitutes a proof of quantum contextuality, since, for a system prepared in the quantum state in which proposition $A$ is true, there is a nonzero probability of finding proposition $B$ or $C$ true and false, respectively. This is, in fact, the method followed in the proofs of quantum contextuality by Stairs \cite{Stairs83}, Clifton \cite{Clifton93a,Clifton93b,Clifton93c}, and Cabello {\em et al.} \cite{CG95,CBTB13}. All these proofs can be then converted into experimental tests of whether or not nature can be described with noncontextual hidden-variable theories \cite{CT11}.

TITSs also serve to prove the KS theorem in any given dimension $d \ge 3$, since, by suitably concatenating several TITSs, one can obtain a set for which noncontextual outcomes satisfying assumptions (i) and (ii) cannot be assigned. Such a set is called a KS set. This is the method followed by Bell \cite{Bell64} and Kochen and Specker \cite{KS67} to prove the KS theorem in $d=3$. The same method can be extended to any $d \ge 3$ \cite{CG96}.

TIFS in which proposition $A$ corresponds to an entangled state and the rest corresponds to product states can be used to prove Bell's theorem (i.e., the impossibility of reproducing QT with local hidden-variable theories). This is exactly what is behind Hardy-like proofs of quantum nonlocality \cite{Hardy93} (for a detailed explanation, see Ref.~\cite{CEG96}).

TITSs are known for any physical system described by a Hilbert space of dimension $d \ge 3$ \cite{KS65, Bell66}. In $d=3$, Bell found one with $n= 13$ propositions \cite{Bell66} and KS found one with $n=10$ \cite{KS67}, which is illustrated in Fig.~\ref{Fig1}(b). Both Bell's and KS's sets belong to a broader family with $n=10+3m$ propositions, with $m=0,1, \ldots$ \cite{CG95}. For $d>3$, TITS with $n=7+d$ are easy to construct from the set of Fig.~\ref{Fig1} by adding the vector with all components zero but the one corresponding to the new dimension \cite{CG95}. However, the problem of which are the {\em minimal} TIFSs and TITSs for any $d \ge 3$ is open. This is the problem we address in this paper.


\section{Method for obtaining minimal TIFSs and TITSs}


A TITS can be also represented by a graph of exclusivity in which $d$-cliques (i.e., $d$ mutually adjacent vertices) represent complete contexts. A graph is said to be \emph{nonrealizable} in dimension $d$ if it represents a set of rays that is not realizable in $S^{d-1},$ i.e., in the unit $(d-1)$ sphere.


{\em Lemma 1} \cite{CDLP13}. The simplest nonrealizable graph of exclusivity in $d=1$ consists of two vertices. The simplest nonrealizable graph of exclusivity in $d=2$ has three vertices with one of them adjacent to the other two. From these to nonrealizable graphs one can recursively construct nonrealizable graphs in any dimension $d$ by starting from the nonrealizable graph in dimension $d-2$ and adding to it two vertices adjacent to all vertices of the nonrealizable graph in $d-2$.


{\em Lemma 2.} Every $n$-vertex graph of exclusivity corresponding to a critical TITS in dimension $d$ contains a ($n+1-d$)-vertex graph of exclusivity corresponding to a TIFS.


{\em Proof.} Let $G$ be a graph of exclusivity corresponding to a TITS in which $A$ true implies $C$ true. Then, every vertex adjacent to $C$ must be false. Then, the induced subgraph of $G$ obtained by removing $C$ and any vertex adjacent both to $A$ and $C$ is a TIFS in which $A$ true implies $B$ false, where $B$ was adjacent to $C$, but not to $A$.\hfill \endproof


{\em Lemma 3.} The graph of exclusivity of a critical TIFS must be biconnected (i.e., it is connected and such that, when removing any vertex, the resulting graph remains connected).


{\em Proof.} Suppose that it is not biconnected. Then, there is, at least, one vertex such that, after removing it, the resulting graph has two unconnected components. If the true and false vertices are in the same component, then this component is a TIFS and, therefore, the original graph of exclusivity is not critical. If the true and false vertices are in different components, then either the removed vertex is false and the component with the true plus the removed vertex form a TIFS, or the removed vertex is not false and the component with the false plus the removed vertex form a TIFS. In both cases, the original graph of exclusivity is not critical.\hfill \endproof


{\em Corollary 1.} Every vertex of a graph of exclusivity corresponding to a TIFS must be adjacent to, at least, two other vertices (i.e., the graph must have minimal valency two).


{\em Lemma 4.} Every graph of exclusivity corresponding to a TIFS in dimension $d$ contains, at least, two $d$-cliques (each of them represented by a $d$-vertex complete graph, i.e., a graph in which all vertices are adjacent).


{\em Proof.} Let $G$ be a graph of exclusivity corresponding to a TIFS and $A$ and $B$ the true and false vertices, respectively. There must be other true vertices $X1,\dots,X_p$. Let be $W=V(G)-\{A,B,X1,\dots,X_p\}$, where $V(G)$ is the set of vertices of $G$. We consider two cases. (a) Every vertex in $W$ belong to the set of vertices of $G$ that are adjacent to $A$, denoted $N(A)$, and $B$ and some $X_i$ are adjacent. In this case, $A \cup N(A)$ and $b \cup N(b)$ form two complete sets. (b) Not all vertices in $W$ belong to $N(A)$. Then, all the vertices in $W$ are false. However, these false vertices are not adjacent to $A$, so their value false must be forced by some other true vertex $X_j$. This vertex is not adjacent to $A$, so it has to belong to a $d$-clique.\hfill \endproof


\section{Dimension 3. Specker's ``bug''}

\begin{figure}
	\begin{center}
		\setlength{\tabcolsep}{1em}
		\begin{tabular}{ccccc}
			\includegraphics{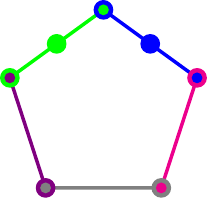}
			&
			\includegraphics{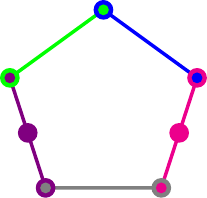}
			\\ 
			\includegraphics{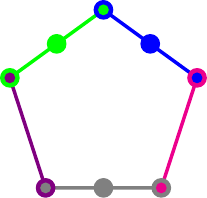}
			& 
			\includegraphics{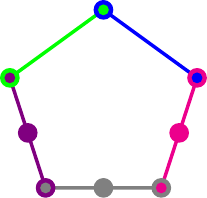}
			\\
		\end{tabular}
		\begin{tabular}{cccccc}
			\includegraphics{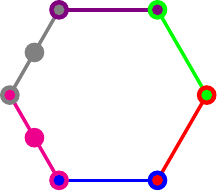}
			&
			\includegraphics{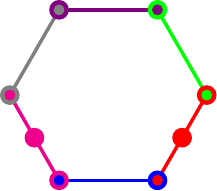}
			&
			\includegraphics{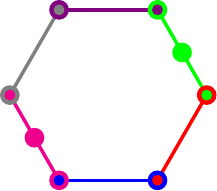}
			\\
			\includegraphics{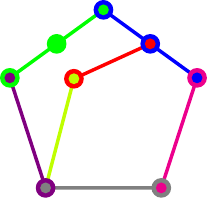} 
			&
			\includegraphics{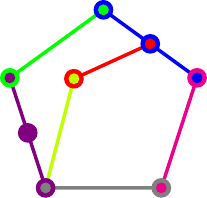} 
			&
			\includegraphics{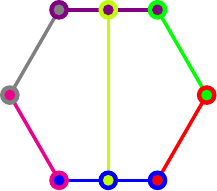} 
		\end{tabular}
	\end{center}
	\centering
	\caption{\label{Fig2}
	(Color online) Greechie orthogonality diagrams of all nonisomorphic 7-vertex (first row) and 8-vertex (the remaining rows)
		biconnected graphs of minimal valence two, not containing cycles of length four, and containing at least two triangles.}
\end{figure}

To obtain the minimal TITS in $d=3$, we combine the previous results as follows:

{\em Step 1.} We generate all nonisomorphic $n$-vertex biconnected graphs (Lemma 3) of minimal valence two (Corollary 1), not containing cycles on length four (Lemma 1), and containing at least two triangles (Lemma 4), with $n \leq 8$. This can be efficiently done using the computer program \texttt{nauty} \cite{McKay07}. We obtain that there are two such graphs for $n=7$ and eight graphs for $n=8$. Their corresponding Greechie orthogonality diagrams are shown in Fig.~\ref{Fig2}.


{\em Step 2.} For every graph obtained after step 1, consider all possible pairs of vertices $(v_i,v_j)$. If, for one $(v_i,v_j)$, the graph does not admit a noncontextual assignment when $v_i=1$ and $v_j=1$ (i.e., then both are true), then the graph is the graph of exclusivity a TIFS in which $A=v_i$ and $B=v_j$. The test of whether or not a graph admits a noncontextual assignment can be done using a simple computer program (e.g., Ref.\ \cite{Peres93}).

After step~2, we find that only the last graph in Fig.~\ref{Fig2} corresponds to a TIFS. This graph, also depicted in Fig.~\ref{Fig1}(a), was first introduced by Kochen and Specker~[2, Fig.~1, p.~182], and later used as a subgraph of the graph $\Gamma_1$ of Kochen and Specker~\cite{KS67}, as depicted in Fig.~\ref{Fig1}(b). Specker referred to this graph as the ``bug.'' This proves that, in $d=3$, there is no TIFS with a smaller number of propositions than the one introduced by Kochen and Specker in 1965 and used in the proofs of Stairs \cite{Stairs83}, Clifton \cite{Clifton93a,Clifton93b,Clifton93c}, and in the simplest Hardy-like proof of quantum contextuality \cite{CBTB13}. This implies, by Lemma 2, that there is no TITS with a smaller number of propositions than the one whose Greechie orthogonality is shown in Fig.~\ref{Fig1}(b).


Orthogonal representations of the minimal TIFS and TITS are presented in the caption of Fig.~\ref{Fig1}. An orthogonal representation of a graph is a set of unit vectors in one-to-one correspondence with the vertices of the graph and such that adjacent vertices are associated orthogonal vectors. It can be easily shown that the minimum angle between the vectors corresponding to vertices $A$ and $B$ is $\arccos\left(\frac{1}{3}\right)$ \cite{Cabello94,Cabello96}. It is interesting to notice that the orthogonal representation of one of the pentagons determines univocally the orthogonal representation of the graph of exclusivity corresponding to a TIFS. This can be seen as follows: suppose we have the vectors corresponding to $A,v_1,v_2,v_3$ and $v_4$. Then, $v_5$ is the vector product of 
$v_1$ and $v_3$. Similarly, $v_6$ is the vector product of 
$v_2$ and $v_4$, and $B$ is the vector product of 
$v_5$ and $v_6$. Also $v_7$ is the vector product of 
$A$ and $B$, and $C$ is the vector product of 
$B$ and $v_7$. Notice also that three nonconsecutive vertices of the pentagon univocally determine the orthogonal representation of the graph of exclusivity.

\begin{figure}
	\centerline{\includegraphics[width=0.78\columnwidth]{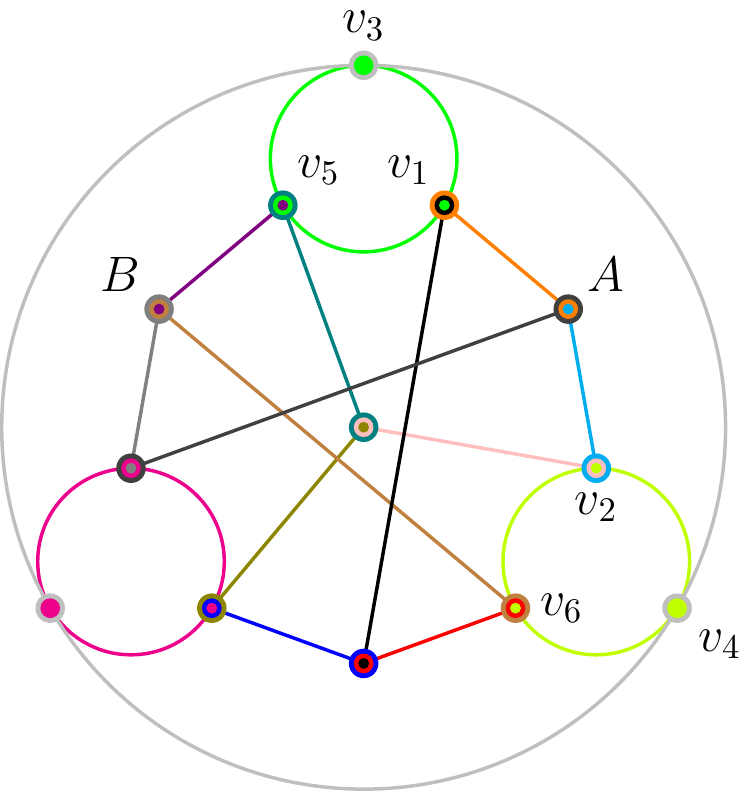}}
	\caption{\label{FigYuOh} (Color online) Greechie orthogonality diagram of the simplest quantum state-independent contextuality set in $d=3$ (and in any $d$) \cite{CKP16}, the Yu-Oh set \cite{YO12}. It contains six TIFSs like the one in Fig.\ \ref{Fig1}(a). One of them is indicated using the same notation used in Fig.\ \ref{Fig1}(a).}
\end{figure}

The state-independent contextuality set with the smallest number of atomic propositions in $d=3$ (and in any dimension $d$ \cite{CKP16}), the Yu-Oh set \cite{YO12}, contains six TIFSs like the one in Fig.~\ref{Fig1}(a). This is shown in Fig.~\ref{FigYuOh}.


\section{Dimension 4. The TIFS in Hardy's proof and other related TIFSs}

\begin{figure}[hbtp]
		\includegraphics[width=\columnwidth]{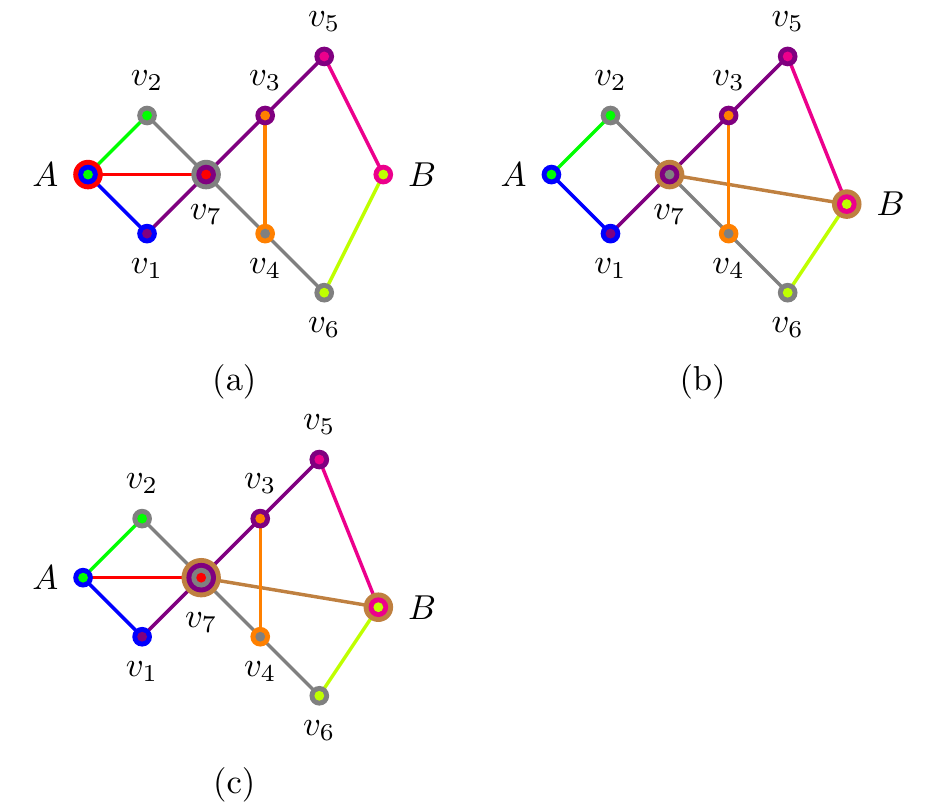} 
	\caption{\label{fig:d4}
		(Color online)
		Greechie orthogonality diagrams of the three minimal
		TIFS in $d=4$. The one in (a) appears 
		in Hardy's proof of Bell's theorem~\cite{Hardy93} (see details in Refs.\ \cite{CEG96,Cabello96,BBC11}).}
\end{figure}

As in the previous section, after an exhaustive computer search, 
we have obtained that there are only three TIFS in $d=4$ with a minimum number of propositions, nine. Their Greechie orthogonality diagrams are depicted in Fig.~\ref{fig:d4}. All three are realizable in $S^3$ by taking, e.g.,
		for Fig.~\ref{fig:d4}(a),
		$A = ({0,-1,\sqrt{2},0} )/\sqrt{3}$,
		$v_1 = ({1,\sqrt{2},1,0} )/ 2 $,
		$v_2 = ({1,0,0,0} )$,
		$v_3 = ({1,0,-1,0} )/\sqrt{2}$,
		$v_4 = ({0,1,0,0} ) $,
		$v_5 = ({-1,\sqrt{2},-1,0} )/ 2$,
		$v_6 = ({0,0,1,0} ) $,
		$v_7 = ({0,0,0,1} )$,
		$B = (\sqrt{1-\varepsilon^2} / \sqrt{3}) ({\sqrt{2},1,0,0})+\varepsilon({0,0,0,1})$;
		for Fig.~\ref{fig:d4}(b),
		$A = (\sqrt{1-\varepsilon^2} / \sqrt{3})({0,-1,\sqrt{2},0} )+\varepsilon({0,0,0,1})$,
		$B = ({\sqrt{2},1,0,0} )/\sqrt{3}$, and $v_i$ as for Fig.~\ref{fig:d4}(a);
		for Fig.~\ref{fig:d4} (c),
		$A = ({0,-1,\sqrt{2},0} )/\sqrt{3}$,
		$B = ({\sqrt{2},1,0,0} ) /\sqrt{3}$, and $v_i$ as for Fig.~\ref{fig:d4}(a).
For the three graphs of exclusivity, the orthogonal representations in $d=4$ are almost unique (except for an $\varepsilon$ value). 
It can be easily shown that the minimum angle between the vectors corresponding to $A$ and $B$ is $\arccos\left(\frac{1-\varepsilon^2}{3}\right)$ 
for the cases in Figs.~\ref{fig:d4}(a) and~\ref{fig:d4}(b) and $\arccos\left(\frac{1}{3}\right)$ for the case in Fig.~\ref{fig:d4}(c).


\section{Minimal TIFSs and TITSs in higher dimensions}


{\em Theorem 2.} Let be $G$ the graph of exclusivity corresponding to a minimal TIFS in dimension $d$; then $|V(G)| = d+5$.


{\em Proof.} First, we prove $|V(G)| \geq d+5$ by induction. For dimensions $ 3 $ and $ 4 $, the theorem has been proven in previous sections. For $ d> 4 $, we know, from Lemma 4, that every graph of exclusivity corresponding to a TIFS in dimension $ d $ contains, at least, two $d$-cliques.
Suppose there is a graph of exclusivity corresponding to a TIFS in dimension $ d $ with less than $ d + 5 $ vertices. It is easy to verify that it cannot contain three $d$-cliques, since this would imply the existence of graphs which are forbidden in dimension $ d $ (see Lemma 1). Then, the two unique $d$-cliques of the graph must have a common vertex. By removing that common vertex, we obtain a graph of exclusivity corresponding to a TIFS in dimension $ d-1 $ with $ | V (G) | = d + 4 $, which contradicts the hypothesis of induction. 

\begin{figure}
	\begin{center}
		\includegraphics[width=\columnwidth]{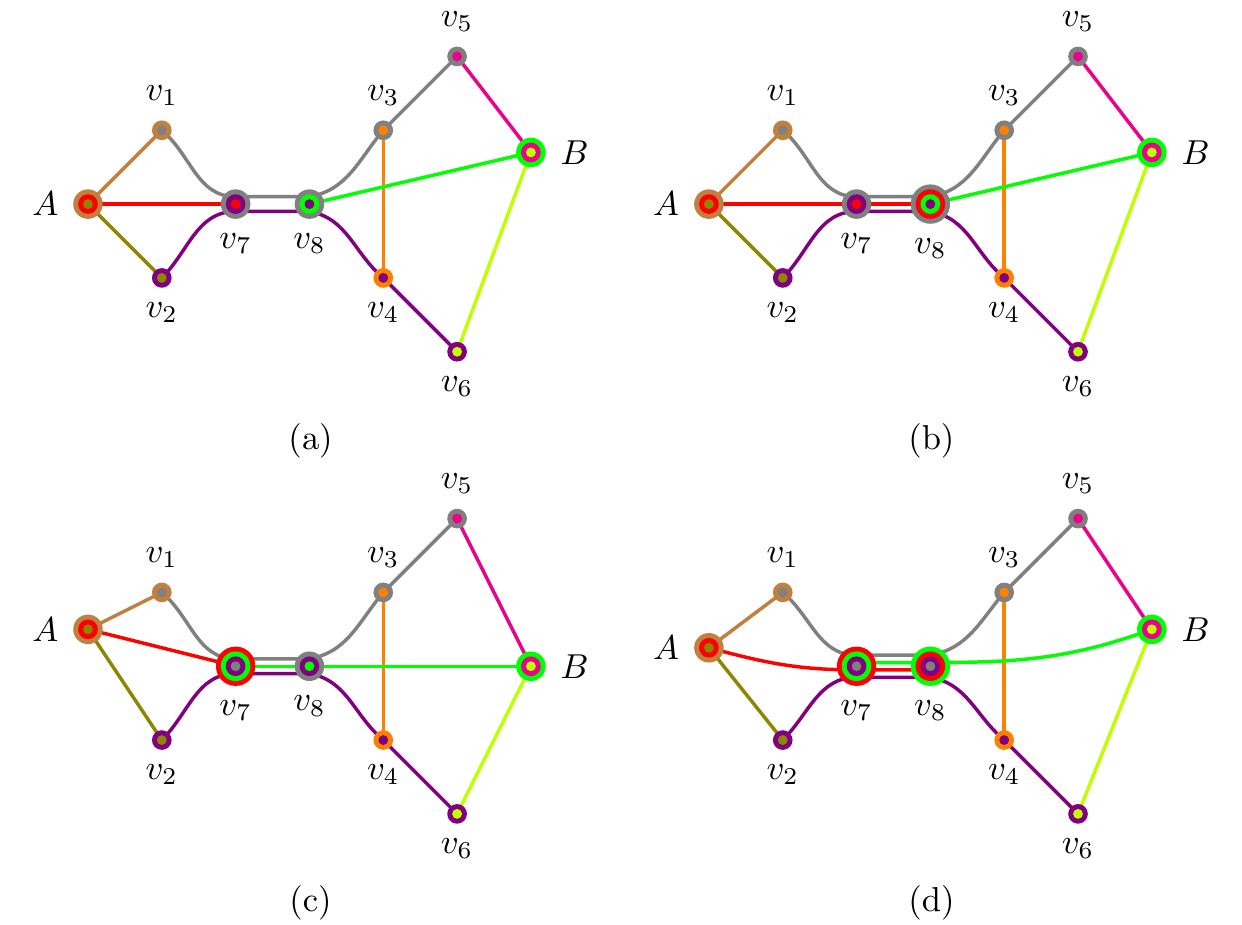}
	\end{center}
	\caption{\label{fig:d5}
		(Color online)
		Greechie orthogonality diagrams of the four minimal
		TIFS in $d=5$. All of them have 10 propositions and nine contexts.
	}
\end{figure}

In order to prove that $|V(G)| \leq d+5$ it suffices to find minimal TIFSs with $ | V (G) | = d + 5 $. Let $G$ be a graph of exclusivity corresponding to a minimal TIFS in dimension $ d> 4 $.
Applying induction, we can verify that $ G $ can be obtained from some graph of exclusivity $ H $ corresponding to a minimal TIFS in dimension $ d-1 $ to adding a vertex $ v $ adjacent to the vertices of the two $d$-cliques of $ H $. The vertex $ v $ may or may not be adjacent to the vertices $ A $ (true) and $ B $ (false), but $ A $ and $ B $ must be each of them adjacent to at least one of the vertices common to the two $d$-cliques of $ H $, because if this condition were not given one of the common vertices to both $d$-cliques of $ G $ could take the value true and $ B $ could take it too. See Fig.~\ref{fig:d5} and Fig~\ref{fig:d6} with examples in dimensions $ 5 $ and $ 6 $. 

For each dimension, the graphs obtained with this method such that the new vertices are always adjacent to $A$ and $B$  are realizable in dimension $d$ by taking, e.g.,  
$A = ({0,-1,\sqrt{2},0,\dots,0} )/\sqrt{3}$,
$v_1 = ({1,\sqrt{2},1,0,\dots,0} )/ 2 $,
$v_2 = ({1,0,0,0,\dots,0} )$,
$v_3 = ({1,0,-1,0,\dots,0} )/\sqrt{2}$,
$v_4 = ({0,1,0,0,\dots,0} ) $,
$v_5 = ({-1,\sqrt{2},-1,0,\dots,0} )/ 2$,
$v_6 = ({0,0,1,0,\dots,0} ) $,
$B = ({\sqrt{2},1,0,0,\dots,0} ) /\sqrt{3}$, 
$v_7 = ({0,0,0,1,\dots,0} )$, \dots,
$v_{d+3} = ({0,0,0,0,\dots,1} )$.\hfill \endproof


Due to this construction, note the following: (i) Orthogonal representations are almost (except for $ \varepsilon $ value) unique for all the graphs of exclusivity corresponding to minimal TIFSs in the same dimension. (ii) The minimum angle for all the graphs of exclusivity corresponding to minimal TIFSs is larger than or equal to $ \arccos \left(\frac{1}{3}\right)$, being able to approach this bound everything we want and being the value achievable when the common vertices to the two $d$-cliques are all adjacent simultaneously to $ A $ and $ B $. We will give an explicit orthogonal representation for all the graphs of exclusivity corresponding to minimal TIFSs of dimensions $ 5 $ and $ 6 $ at the end of this section. The general construction for any dimension is immediate from these examples.


As all the graphs of exclusivity corresponding to minimal TIFSs can be found by a constructive method, we can count them. The number of minimal TIFSs in dimension $ 3,4,5,6,7,8, \ldots$ is $ 1,3,4,8,13,19, \ldots $. The number of minimal TIFSs in dimension $ d = 3,4 $ is $ (d-1) (d-2) / 2 $. In higher dimensions ($ d \geq 5 $) the number of minimal TIFS is $ \frac {(d-1) (d-2)} {2} -2 $.
To count the number of graphs of exclusivity corresponding to minimal TITSs, note that the vertices added in the construction form a $(d-3)$-clique. It suffices then to count the possible connections (except isomorphisms) between the vertices of the $(d-3)$-clique and the vertices $A$ (true) and $B$ (false). Each vertex of the $(d-3)$-clique can have three different and incompatible states (adjacent to true, adjacent to false or adjacent to both); we have combinations with repetition of three elements taken in groups of $ d-3 $. This provides $\mathrm{CR}^{d-3}_3 = \binom{d-1}{d-3} = \binom{d-1}{2} = (d-1) (d-2) / 2 $, where $\mathrm{CR}$ stands for combinations with repetition. If the dimension is larger than $ 4 $, it is necessary to eliminate the graphs where all vertices of the $(d-3)$-clique are adjacent only to $ A $ or only to $ B $ obtaining $ \frac{(d-1) (d-2)}{2} -2 $.


{\em Theorem 3.} Minimal TITSs have $d+7$ propositions in dimension $d\geq 3$.

\begin{figure}
\centerline{\includegraphics[width=\columnwidth]{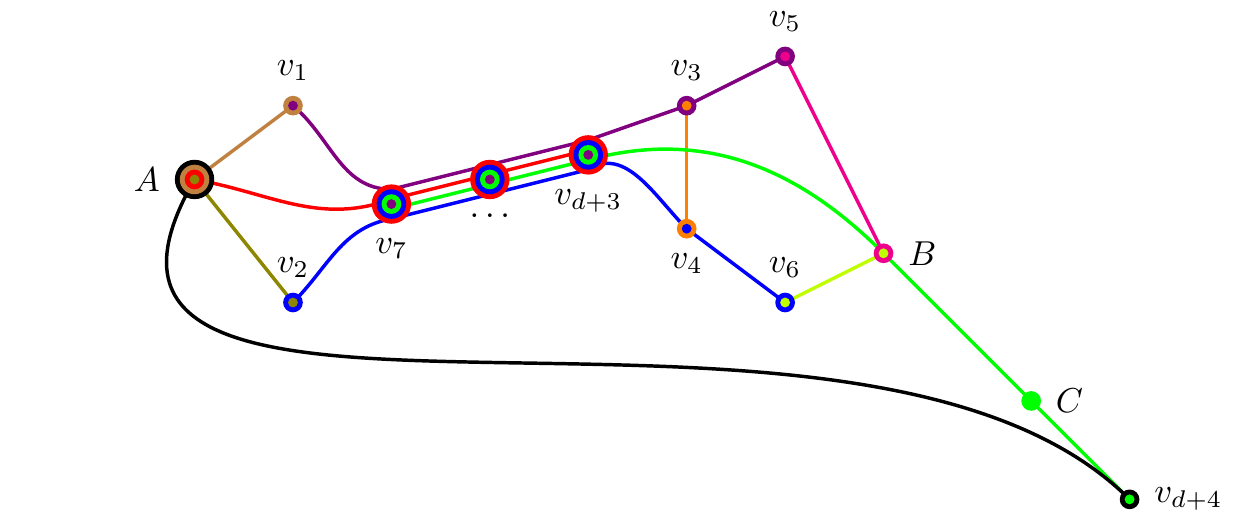}}
\caption{\label{fig:scheme}(Color online) Scheme for constructing a minimal TITS with $d+7$ propositions in dimension $d$. The subgraph $\{A,B,v_1,\dots,v_{d+5},B\}$ corresponds to a minimal TIFS. If $A$ is true then $B$ is false and also $v_7,\dots,v_{d+3},$ and $v_{d+4}$ are false. Therefore, since $\{v_7,\dots,v_{d+3},v_{d+4},B,C\}$ are mutually exclusive, then $C$ must be true.}
\end{figure}

{\em Proof.} Suppose that the minimal TITS has less than $d+7$ propositions and that $A$ true implies $C$ true. Therefore, the true of $C$ is forced by a $d$-clique. Then, at least one of the vertices of this $d$-clique, say vertex $u$, is not adjacent to $A$. Otherwise, a forbidden subgraph would appear. Therefore, we can remove the vertex $C$ and all the vertices of the $d$-clique, except $u$, and construct a TIFS ($A$ true implies $u$ false) with less than $d+5$ vertices, and this is in contradiction with Theorem 2. On the other hand, the addition of two vertices to the graph of exclusivity corresponding to the minimal TIFS in dimension $d$, as shown in Fig.~\ref{fig:scheme}, provides a minimal TITS with $d+7$ propositions in dimension $d\geq 3$. \hfill \endproof


{\em Corollary 2.} The graphs of exclusivity corresponding to the minimal TITSs with $d+7$ propositions contains exactly three $d$-cliques.


\subsection{Dimension 5}


We have obtained that there are four TIFSs with a minimum number of propositions in $d=5$. Their Greechie orthogonality diagrams are shown in Fig.~\ref{fig:d5}.
These TIFSs are realizable in $S^4$ by taking,
		e.g., for Fig.~\ref{fig:d5}(a),
		$A = (\sqrt{1-\varepsilon^2} / \sqrt{3})({0,-1,\sqrt{2},0,0} )+\varepsilon({0,0,0,0,1})$,
		$v_1 = ({1,\sqrt{2},1,0,0} )/ 2 $,
		$v_2 = ({1,0,0,0,0} )$,
		$v_3 = ({1,0,-1,0,0} )/\sqrt{2}$,
		$v_4 = ({0,1,0,0,0} ) $,
		$v_5 = ({-1,\sqrt{2},-1,0,0} )/ 2$,
		$v_6 = ({0,0,1,0,0} ) $,
		$v_7 = ({0,0,0,1,0} )$,
		$v_{8} = ({0,0,0,0,1} )$, and
		$B = (\sqrt{1-\varepsilon^2} / \sqrt{3}) ({\sqrt{2},1,0,0,0})+\varepsilon({0,0,0,1,0})$; for Fig.~\ref{fig:d5}(b),
		$A = ({0,-1,\sqrt{2},0,0,} )/\sqrt{3}$,
		$B = (\sqrt{1-\varepsilon^2} / \sqrt{3}) ({\sqrt{2},1,0,0,0})+\varepsilon({0,0,0,1,0})$,
		and $v_i$ as for Fig.~\ref{fig:d5}(a); for Fig.~\ref{fig:d5}(c),
		$A = (\sqrt{1-\varepsilon^2} / \sqrt{3})({0,-1,\sqrt{2},0,0} )+\varepsilon({0,0,0,1,0})$,
		$B = ({\sqrt{2},1,0,0,0} )/\sqrt{3}$, 
		and $v_i$ as for Fig.~\ref{fig:d5}(a); for Fig.~\ref{fig:d5}(d),
		$A = ({0,-1,\sqrt{2},0,0} )/\sqrt{3}$,
		$B = ({\sqrt{2},1,0,0,0} )/\sqrt{3}$, 
		and $v_i$ as for Fig.~\ref{fig:d5}(a).
		Notice that the orthogonal representations are almost unique (except for an $\varepsilon$ value). 
The minimum angle between the vectors corresponding to $A$ and $B$ is $\arccos\left(\frac{1-\varepsilon^2}{3}\right)$ for all the cases in Fig.~\ref{fig:d5} except for case (d), that is $\arccos\left(\frac{1}{3}\right)$. These realizations admit many implementations in QT, depending on the physical meaning of the canonical basis.


\subsection{Dimension 6}

\begin{figure*}
	\includegraphics{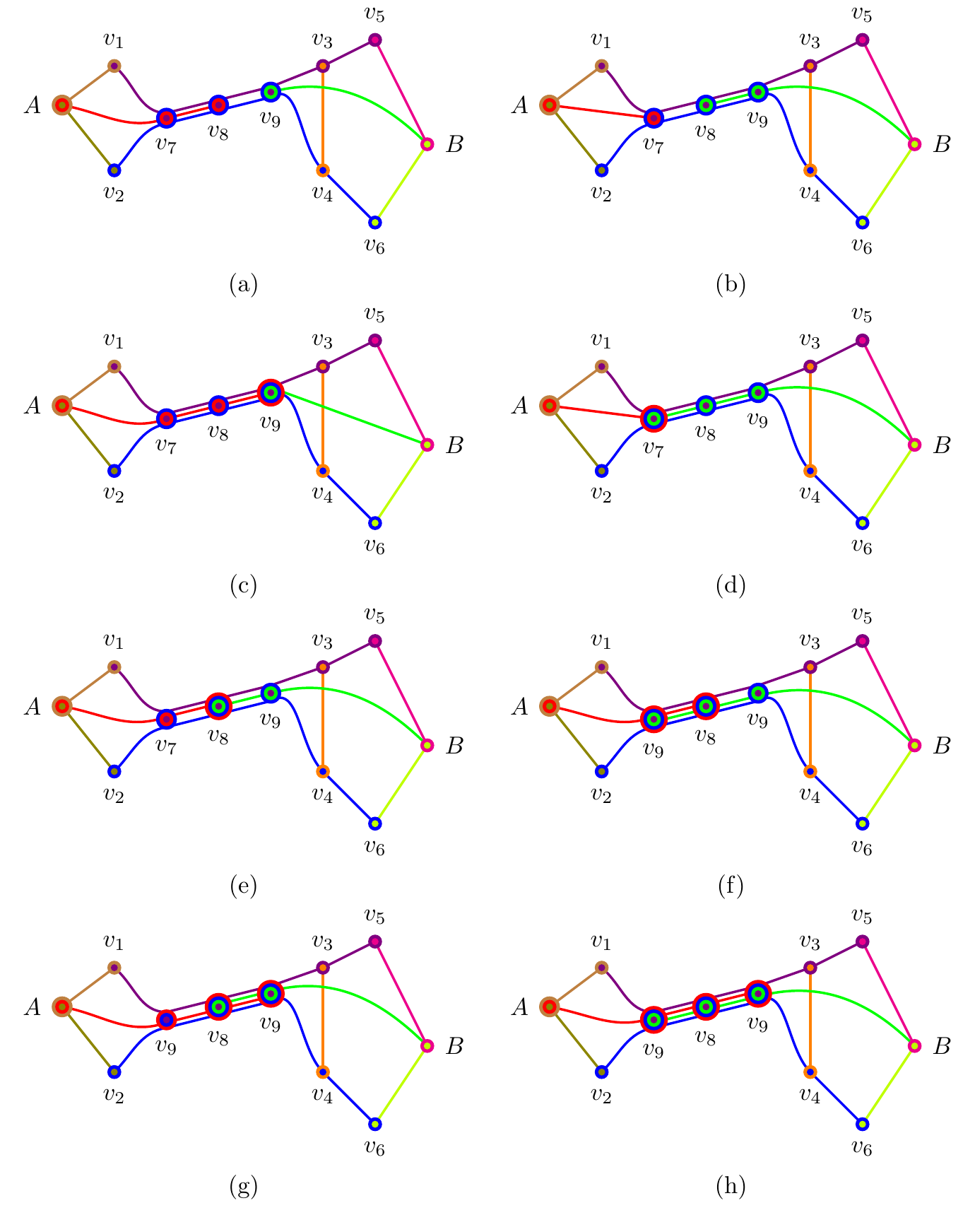}
	\caption{\label{fig:d6}
		(Color online)
		Greechie orthogonality diagrams of the eight minimal TIFs in $d=6$. All of them have 11 propositions and nine contexts.}
\end{figure*}

We have obtained that there are eight TIFS with a minimum number of propositions in $d=6$. Their Greechie orthogonality diagrams are shown in Fig.~\ref{fig:d6}. All of them are realizable in $S^5$ by taking, e.g., for Fig.~\ref{fig:d6}(a),
$A = (\sqrt{1-\varepsilon^2} / \sqrt{3})({0,-1,\sqrt{2},0,0,0} )+\varepsilon({0,0,0,0,0.1})$,
$v_1 = ({1,\sqrt{2},1,0,0,0} )/ 2 $,
$v_2 = ({1,0,0,0,0,0} )$,
$v_3 = ({1,0,-1,0,0,0} )/\sqrt{2}$,
$v_4 = ({0,1,0,0,0,0} ) $,
$v_5 = ({-1,\sqrt{2},-1,0,0,0} )/ 2$,
$v_6 = ({0,0,1,0,0,0} ) $,
$v_7 = ({0,0,0,1,0,0} )$,
$v_{8} = ({0,0,0,0,1,0} )$,
$v_{9} = ({0,0,0,0,0,1} )$, and
$B = (\sqrt{1-\varepsilon^2} / \sqrt{3}) ({\sqrt{2},1,0,0,0,0})+(\varepsilon/\sqrt{2})({0,0,0,1,1,0})$;
for Fig.~\ref{fig:d6}(b),
$A = (\sqrt{1-\varepsilon^2} / \sqrt{3}) ({0,-1,\sqrt{2},0,0,0})+(\varepsilon/\sqrt{2})({0,0,0,0,1,1})$,
$B = (\sqrt{1-\varepsilon^2} / \sqrt{3}) ({\sqrt{2},1,0,0,0,0})+\varepsilon({0,0,0,1,0,0})$, 
		and the remaining $v_i$ as for Fig.~\ref{fig:d6}(a); 
for Fig.~\ref{fig:d6}(c),
$A = ({0,-1,\sqrt{2},0,0,0})/ \sqrt{3}$,
$B = (\sqrt{1-\varepsilon^2} / \sqrt{3}) ({\sqrt{2},1,0,0,0,0})+(\varepsilon/\sqrt{2})({0,0,0,1,1,0})$,
		and the remaining $v_i$ as for Fig.~\ref{fig:d6}(a);
for Fig.~\ref{fig:d6}(d),
$A = (\sqrt{1-\varepsilon^2} / \sqrt{3}) ({0,-1,\sqrt{2},0,0,0})+(\varepsilon/\sqrt{2})({0,0,0,0,1,1})$,
$B = ({\sqrt{2},1,0,0,0,0})/ \sqrt{3}$,
		and $v_i$ as for Fig.~\ref{fig:d6}(a);
for Fig.~\ref{fig:d6}(e),
$A = (\sqrt{1-\varepsilon^2} / \sqrt{3}) ({0,-1,\sqrt{2},0,0,0})+\varepsilon({0,0,0,0,0,1})$,
$B = (\sqrt{1-\varepsilon^2} / \sqrt{3}) ({\sqrt{2},1,0,0,0,0})+\varepsilon({0,0,0,1,0,0})$,
		and $v_i$ as for Fig.~\ref{fig:d6}(a);
for Fig.~\ref{fig:d6}(f),
$A = (\sqrt{1-\varepsilon^2} / \sqrt{3}) ({0,-1,\sqrt{2},0,0,0})+\varepsilon({0,0,0,0,0,1})$,
$B = ({\sqrt{2},1,0,0,0,0})/ \sqrt{3}$,
		and $v_i$ as in for Fig.~\ref{fig:d6}(a);
for Fig.~\ref{fig:d6}(g),
$A = ({0,-1,\sqrt{2},0,0,0})/ \sqrt{3}$,
$B = (\sqrt{1-\varepsilon^2} / \sqrt{3}) ({\sqrt{2},1,0,0,0,0})+\varepsilon({0,0,0,1,0,0})$,
		and $v_i$ as for Fig.~\ref{fig:d6}(a);
for Fig.~\ref{fig:d6}(h),
$A = ({0,-1,\sqrt{2},0,0,0})/ \sqrt{3}$,
$B = ({\sqrt{2},1,0,0,0,0})/ \sqrt{3}$,
		and $v_i$ as for Fig.~\ref{fig:d6} (a).
		Notice that the orthogonal representations are almost unique (except for an $\varepsilon$ value).
The minimum angle between the vectors corresponding to $A$ and $B$ is $\arccos\left(\frac{1-\varepsilon^2}{3}\right)$ for all cases in Fig.~\ref{fig:d6} except for (h), that is $\arccos\left(\frac{1}{3}\right)$.


\section{Open problems}


Here we have identified the simplest TIFSs and TITSs in every finite dimension. TIFSs and TITSs are not only important for themselves, but also because they are related to some open problems. For example, in Ref.\ \cite{Peres03} Peres conjectured that the KS set with the smallest number of atomic propositions in any dimension is the one in Ref.\ \cite{CEG96}, with 18 propositions in $d=4$. The intermediate results we have developed in this paper can help to prove this conjecture. Another open problem that can benefit from our results is identifying the KS set in $d=3$ with the smallest set of atomic propositions. Curiously, after more than $50$ years, this problem remains open.

Other interesting open problem is identifying the minimal true-iff-true sets in every finite dimension $d$. A true-iff-true set (also called nonseparating set) is one that contains two propositions which must be both true or both false. This is not the same as in a TITS, as for a TITS, $C$ true does not imply $A$ true. For $d=3$, a true-iff-true-set was identified in Ref.\ \cite{KS67}. These sets are interesting because they demonstrate an even larger conflict between QT and noncontextual hidden-variable theories as, although there still exist classical valuations and truth tables, they are more in contradiction with QT, up to the point where propositional structures containing these sets cannot be embedded into any kind of hidden parameter model \cite{KS67}, such as partition logics~\cite{svozil-2001-eua}, and their model realizations as Wright's generalized urn model~\cite{wright}, or automaton logic~\cite{schaller-96} (still allowing logics with TIFS or TITS). We conjecture that the 17-ray true-iff-true set in Ref.\ \cite{KS67} is minimal in $d=3$. However, we do not have a proof.


\begin{acknowledgments}
 AC acknowledges support from Project No.\ FIS2017-89609-P, ``Quantum Tools for Information, Computation and Research'' (MINECO, Spain) with FEDER funds, the FQXi Large Grant ``The Observer Observed: A Bayesian Route to the Reconstruction of Quantum Theory,'' and the Project ``Photonic Quantum Information'' (Knut and Alice Wallenberg Foundation, Sweden).
\end{acknowledgments}







\end{document}